\documentclass[aps, prb, reprint, superscriptaddress]{revtex4-1}
\usepackage{amssymb}
\usepackage{amsmath}
\usepackage{graphicx}
\usepackage{bm}
\usepackage{color}

\input{tcilatex}
\begin{document}

\title{Non-thermal Tomonaga-Luttinger liquid eventually emerging from hot
electrons in the quantum Hall regime}
\author{Kotaro Suzuki}
\author{Tokuro Hata}
\author{Yuya Sato}
\affiliation{Department of Physics, Tokyo Institute of Technology, 2-12-1 Ookayama,
Meguro, Tokyo, 152-8551, Japan.}
\author{Takafumi Akiho}
\author{Koji Muraki}
\affiliation{NTT Basic Research Laboratories, NTT Corporation, 3-1 Morinosato-Wakamiya,
Atsugi 243-0198, Japan.}
\author{Toshimasa Fujisawa}
\email{fujisawa@phys.titech.ac.jp}
\affiliation{Department of Physics, Tokyo Institute of Technology, 2-12-1 Ookayama,
Meguro, Tokyo, 152-8551, Japan.}
\date{\today }
\maketitle

\textbf{Dynamics of integrable systems, such as Tomonaga-Luttinger (TL)
liquids, is deterministic, and the absence of stochastic thermalization
processes provides unique characteristics, such as long-lived non-thermal
metastable states with many conserved quantities. Here, we show such
non-thermal states can emerge even when the TL liquid is excited with
extremely high-energy hot electrons in chiral quantum-Hall edge channels.
This demonstrates the robustness of the integrable model against the
excitation energy. Crossover from the single-particle hot electrons to the
many-body TL liquid is investigated by using on-chip detectors with a
quantum point contact and a quantum dot. The charge dynamics can be
understood with a single-particle picture only for hot electrons. The
resulting electron-hole plasma in the TL liquid shows a non-thermal
metastable state, in which warm and cold electrons coexist without further
thermalization. The multi-temperature constituents are attractive for
transporting information with conserved quantities along the channels.}

\bigskip

\textbf{Introduction}

The Coulomb interaction in one-dimensional (1D) conductors plays an
essential role in non-equilibrium transport characteristics\cite%
{BookGiamarchi,ReviewElecQO,TaruchaSSC1995,BockrathNature1999,LorenzNature2002,AuslaenderScience2005}%
. Two distinct transport regimes, Tomonaga-Luttinger (TL) liquid in the
low-energy regime and single-particle hot-electron transport in the
high-energy regime, appear in chiral 1D channels of an integer quantum Hall
(QH) region at Landau-level filling factor $\nu $ = 2 in AlGaAs/GaAs
heterostructures\cite{BookEzawa,ChangRMB,LevkivskyiBook2012}. In the
low-energy regime, the energy dispersion near the Fermi energy can be
approximated to be linear and the system is well described with collective
excitation modes in the TL liquid model\cite{vonDelftAnnPhys1998,BergPRL2009}%
. Arbitrary electronic excitation can be represented by spin and charge
collective modes. The collective excitations propagate at different
velocities for each mode, a phenomenon known as spin-charge separation\cite%
{FreulonNatComm-SC,Bocquillon-NatCom2013,Inoue-PRL2014,Kamata-NatNano2014,Hashisaka-NatPhys2017}%
. Owing to the integrable nature of the TL model, the electronic excitation
will be equilibrated into an ensemble of spin and charge excitations but not
fully relaxed into a thermal equilibrium state\cite%
{GutmanPRB2010,GutmanPRL2008,Iucci2009,KovrizhinPRB2011,LevkivskyiPRB2012}.
This non-thermal metastable state (prethermalized state) survives even after
traveling over a distance (\TEXTsymbol{>} 20 $\mu $m) much longer than
required for the spin-charge separation (\TEXTsymbol{<} 0.1 $\mu $m)\cite%
{WashioPRB2016,Itoh-PRL2018,RodriguezNatComm2020}. In these experiments, the
initial state is excited with low energies (\TEXTsymbol{<} 0.2 meV from the
Fermi energy) where the TL model holds well. The validity of the TL model,
or the system's integrability, is no longer guaranteed in the high-energy
regime, where due to the deviation from the linear energy dispersion the
Coulomb interaction induces weak electron-electron (e-e) scattering between
the hot and cold electrons\cite{TaubertPRB2011,OtaPRB2019}.

In the high energy regime, the hot electron travels for a long distance (%
\TEXTsymbol{>} 10 $\mu $m) without losing energy when the Coulomb
interaction is suppressed, for example, by screening the interaction with
metal on the surface\cite%
{FletcherPRL2013,KataokaPRL2016,JohnsonPRL2018,FletcherNatCom2019,Ubbelohde-NatNano2014}%
. Such ballistic hot electron transport can be seen at high energy (%
\TEXTsymbol{>} 50 meV) from the Fermi energy. In the intermediate energy
region, the e-e scattering becomes significant, where the crossover between
the TL liquid and single-particle hot electron transport is expected.
However, the crossover regime remained veiled as the cold- and hot-electron
dynamics have been studied separately by using different experimental
schemes.

Here, we investigate the crossover dynamics from high-energy hot electrons
to a non-thermal TL liquid, where the hot electrons lose their energy by
exciting the TL liquid under the Coulomb interaction. The resulting state is
unusual as warm and cold electrons coexist in the same channel even in the
quasi-steady state. The emergence of the non-thermal many-body state is
striking as the hot-electron energy is well beyond the low energy regime for
the TL liquid. The obtained non-thermal state is stable for long transport
even after full relaxation of hot electrons, which represents the robustness
of the integrable nature. We find that the final state depends on the
initial hot-electron energy even after the equilibration, while
energy-dependent dissipation is observed at higher energy. The crossover
includes intra-channel scattering generating a hot spot in the spin-up
channel and subsequent inter-channel scattering generating another hot spot
in the spin-down channel. The characteristics are experimentally obtained by
using a quantum point contact (QPC) as a spin-dependent bolometer to extract
the energy-space trajectory of hot electrons and a quantum dot (QD) as a
narrow-band spectrometer to evaluate the non-thermal TL liquid. The QH
channel is a promising platform to study the interplay between
single-particle and many-body physics.

\begin{figure*}[tbp]
\begin{center}
\includegraphics[width = 6.6 in]{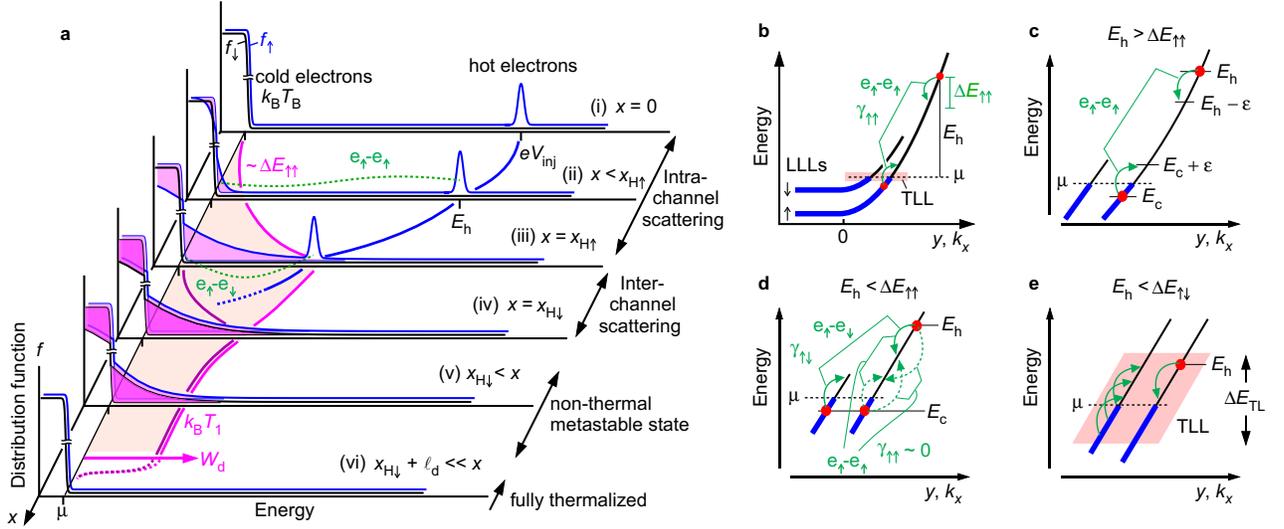}
\end{center}
\caption{\textbf{Crossover dynamics of single-particle scattering processes
and non-thermal Tomonaga-Luttinger (TL) liquid.} \textbf{a} Schematic
spatial evolution of energy distribution functions $f_{\uparrow }$ and $%
f_{\downarrow }$ for up- and down-spins, respectively. Hot electrons are
injected into the spin-up channel at energy $eV_{\mathrm{inj}}$ (i) and
relax with intra-channel electron-electron scattering (e$_{\uparrow }$-e$%
_{\uparrow }$) and inter-channel scattering (e$_{\uparrow }$-e$_{\downarrow
} $) (i)-(iv). Electron-hole plasma is excited in the channels, and hot
spots appear at $x_{\mathrm{H\uparrow }}$ and $x_{\mathrm{H\downarrow }}$ in
the spin-up and -down channels, respectively. The system ends up in a
non-thermal metastable state (v) before reaching thermal equilibrium (vi). 
\textbf{b} Energy profiles for the lowest Landau levels (LLLs) for up and
down spins. Intra-channel electron-electron scattering (e$_{\uparrow }$-e$%
_{\uparrow }$) is shown for a hot electron at energy $E_{\mathrm{h}}$. 
\textbf{c} Intra-channel scattering (e$_{\uparrow }$-e$_{\uparrow }$) for
hot and cold electrons at energies $E_{\mathrm{h}}$ and $E_{\mathrm{c}}$,
respectively, and final states at $E_{\mathrm{h}}-\protect\varepsilon $ and $%
E_{\mathrm{c}}+\protect\varepsilon $ with energy exchange $\protect%
\varepsilon $. \textbf{d} Inter-channel scattering (e$_{\uparrow }$-e$%
_{\downarrow }$) and suppressed intra-channel scattering (e$_{\uparrow }$-e$%
_{\uparrow }$) with destructive interference for $E_{\mathrm{h}}<$ $\Delta
E_{\uparrow \uparrow }$. \textbf{e} Formation of a TL liquid within the
energy range $\Delta E_{\mathrm{TL}}$.}
\end{figure*}

\bigskip

\textbf{Results and discussion}

\textbf{Crossover dynamics.} We investigate the dynamics of hot and cold
electrons in the two chiral edge channels at $\nu $ = 2 in a high magnetic
field $B$. Figure 1a illustrates our finding on how the energy distribution
functions $f_{\uparrow }$ and $f_{\downarrow }$ of up and down spins,
respectively, change with traveling along the $x$-axis. With the spin-up and
down channels equilibrated with cold electrons prepared at the base
temperature $T_{\mathrm{B}}$ at $x<0$, we inject hot electrons with energy $%
eV_{\mathrm{inj}}$ selectively to the spin-up channel at $x=0$ [panel (i)].
For moderate $eV_{\mathrm{inj}}$, the Coulomb interaction induces
intra-channel e-e scattering between the hot and cold spin-up electrons (e$%
_{\uparrow }$-e$_{\uparrow }$), and electron-hole plasma (the light magenta
regions) is generated in the spin-up channel [panel (ii)] \cite{OtaPRB2019}.
We neglect three-body e-e scattering throughout this paper for simplicity%
\cite{KhodasPRB2007}. The hot-electron relaxation can be characterized by
the energy decay rate per unit length, $\gamma _{\uparrow \uparrow }=-dE_{%
\mathrm{h}}/dx$, for hot-electron's energy $E_{\mathrm{h}}$ and the maximum
energy exchange in a single scattering event, $\Delta E_{\uparrow \uparrow }$%
\cite{LundePRB2010-EE,LundePRB-EE}. $\gamma _{\uparrow \uparrow }$ is
obtained by measuring the energy($E_{\mathrm{h}}$)-space($x$) trajectory of
the hot electrons, and $\Delta E_{\uparrow \uparrow }$ is roughly estimated
from the spread of the distribution function, $W_{\mathrm{d}}$, near the
chemical potential $\mu $. As we show later, both $\gamma _{\uparrow
\uparrow }$ and $\Delta E_{\uparrow \uparrow } $ increase rapidly as the hot
electrons lose energy. The resultant sharp drop in the hot electron's energy
creates a hot spot at $x=x_{\mathrm{H\uparrow }}$, where the spin-up
electrons are maximally excited [panel (iii)]. In addition, the
inter-channel e-e scattering between up and down spins (e$_{\uparrow }$-e$%
_{\downarrow }$) excites plasma (the dark magenta regions) and creates
another hot spot at $x=x_{\mathrm{H\downarrow }}$ in the spin-down channel
[panel (iv)]. The e$_{\uparrow }$-e$_{\downarrow }$ scattering equilibrates
the two channels further during transport. Importantly, once the
electron-hole plasma relaxes to fit in the energy range where the TL physics
governs, the plasma cannot equilibrate further and remains in a non-thermal
metastable state in which warm and cold electrons coexist in the same
channel [panel (v)]. Such crossover from hot electrons into a non-thermal TL
liquid is the subject of this paper. Due to the coupling to the environment,
the system finally reaches thermal equilibrium at $T_{\mathrm{B}}$ after
long transport [panel (vi)].

The crossover dynamics can be understood by considering microscopic
processes in the edge states, where the Landau levels are bent upwards by
the edge potential, as schematically shown in Fig. 1b. This energy($E$) -
space($y$) diagram can be understood with the energy($E$) - wavenumber($%
k_{x} $) dispersion, as the relation $k_{x}=-eBy/\hbar $ holds in the QH
regime under the Landau gauge\cite{BookEzawa}. Due to the nonlinearity of
the dispersion, e-e scattering is suppressed as it cannot conserve both
energy and momentum, particularly when the energy $E_{\mathrm{h}}$ and the
energy exchange $\varepsilon $ are large (Fig. 1c). In reality, the
scattering is allowed by the presence of disorder that breaks the
translational invariance and lifts the momentum conservation to some extent.
This should determine the cutoff $\Delta E_{\uparrow \uparrow }$ in the e-e
scattering. For the concave dispersion relation, $\Delta E_{\uparrow
\uparrow }$ should increase with decreasing $E_{\mathrm{h}}$. The Coulomb
interaction increases with decreasing the distance at lower $E_{\mathrm{h}}$%
. Both effects induce the rapid reduction of $E_{\mathrm{h}}$ with large $%
\gamma _{\uparrow \uparrow }$ and $\Delta E_{\uparrow \uparrow }$ near $x=x_{%
\mathrm{H\uparrow }}$.

When $E_{\mathrm{h}}<\Delta E_{\uparrow \uparrow }$ is reached, the system
enters a new regime, where the scattering involves two competing processes
with the swapped final states [the dashed lines with an arrow labeled e$%
_{\uparrow }$-e$_{\uparrow }$ in Fig. 1d]. They destructively interfere with
each other, and thus the intra-channel scattering of the same spins should
be suppressed at low $E_{\mathrm{h}}$ ($<\Delta E_{\uparrow \uparrow }$)\cite%
{LundePRB-EE}. When the intra-channel scattering is inefficient, the
inter-channel scattering between different spins should dominate the
relaxation. This region can be referred to as the inter-channel scattering
regime. The hot electrons in the spin-up channel lose their energies by
exciting an electron-hole plasma in the spin-down channel, and no competing
process is present for scattering between different spins. The inter-channel
scattering subsequently excites the spin-up channel and thus equilibrates
the two channels in the end. As the spin-down channel is located on the
inner side, where the edge potential is more gentle, the maximum energy
exchange $\Delta E_{\uparrow \downarrow }$ and the energy loss rate $\gamma
_{\uparrow \downarrow }$ for different spins should be smaller than $\Delta
E_{\uparrow \uparrow }$ and $\gamma _{\uparrow \uparrow }$, respectively.

The above single-particle picture fails when the hot electron is further
relaxed within the energy range $\Delta E_{\mathrm{TL}}$ of the TL regime ($%
E_{\mathrm{h}}\lesssim \Delta E_{\mathrm{TL}}$). The dispersion can be
approximated to be linear, as shown in Fig. 1e. All electrons within the
energy range $\Delta E_{\mathrm{TL}}$ near the Fermi level interact with
each other. Such interacting electrons can be understood with
non-interacting bosonic excitations in the TL model \cite%
{vonDelftAnnPhys1998,KovrizhinPRB2011}. The inter-channel interaction
induces spin-charge separation, and the intra-channel interaction enhances
the charge velocity.\cite{Hashisaka-NatPhys2017} Once all electron-hole
plasma enters the TL regime, no further transformation is expected in the
plasma. Therefore, a non-thermal metastable state eventually shows up after
stochastic processes. In the following experiment, the coexistence of warm
and cold electrons remains even after long-distance transport. While the TL
model is based on low-energy physics, such non-thermal states can emerge
from high-energy hot electrons. This demonstrates the robustness of the
prethermalized state against the excitation energy.

Electronic distribution functions of non-thermal states have been studied
previously. When an initial state with a double-step distribution function
is prepared by a QPC with bias voltage $V_{\mathrm{b}}$ and low transmission
probability, the non-thermal state shows an arctangent distribution function
in the low-energy limit at $\left\vert E-\mu \right\vert \ll eV_{\mathrm{b}}$
\cite{LevkivskyiPRB2012,Itoh-PRL2018}. Phenomenologically, the non-thermal
states in the high-energy region can be approximated by a binary
distribution function 
\begin{equation}
f\left( E\right) =\left( 1-p\right) f_{\mathrm{FD}}\left( E;k_{\mathrm{B}%
}T_{0}\right) +pf_{\mathrm{FD}}\left( E;k_{\mathrm{B}}T_{1}\right)
\end{equation}%
consisting of high- ($T_{1}$) and low-temperature ($T_{0}$, close to the
base temperature $T_{\mathrm{B}}$) components with Fermi distribution $f_{%
\mathrm{FD}}\left( E;k_{\mathrm{B}}T\right) =1/\left[ 1+\exp \left( E/k_{%
\mathrm{B}}T\right) \right] $, which is valid for small fraction $p$ ($\ll 1$%
) \cite{WashioPRB2016}. The binary form captures the exponential current
profiles $\sim e^{-E/k_{\mathrm{B}}T_{1}}$ observed in the low-energy regime 
\cite{WashioPRB2016,Itoh-PRL2018,RodriguezNatComm2020}. As we do not know
the theoretical form for hot-electron injection, we use this binary form to
analyze the non-thermal states in this paper.

\bigskip

\textbf{Experimental setup.} We investigated the crossover dynamics with the
device fabricated in a standard AlGaAs/GaAs heterostructure with an electron
density\ of 2.5$\times $10$^{11}$ cm$^{-2}$ and low-temperature mobility of
about 10$^{6}$ cm$^{2}$/Vs, as shown in Fig. 2a. The electron system is
divided into three conductive regions that serve as the source (S), base
(B), and drain (D) by applying large negative gate voltages $V_{\mathrm{G}i}$
on gate $i=1,2,3,\cdots $. A perpendicular magnetic field $B=$ 5.0 T is
applied to set the electron system in the $\nu =2$ QH regime and form two
parallel edge channels (the blue lines) in each region.

Hot electrons with up spin and energy $eV_{\mathrm{inj}}$ ($=$ 1 $-$ 100
meV) are injected from S to B, by applying voltage $-V_{\mathrm{inj}}$ to
the ohmic contact of S and tuning the transmission of the injector point
contact (IPC). The electronic distribution function of the up-spin channel
near the chemical potential is investigated with a QD, and electron-hole
plasma in up- and down-spin channel are studied by activating a QPC as shown
in Fig. 2b. The QD or QPC is located at a distance $L$ from the IPC, where
several injection gates at different distances $L=$ 1, 2, 5, 10, 20, and 30 $%
\mu $m were selectively activated [see the scanning electron micrograph of
Fig. 2c]. As shown in the schematic energy profile of Fig. 2d, the hot spot (%
$x_{\mathrm{H\uparrow }}$) can be changed by tuning $eV_{\mathrm{inj}}$. The
non-thermal state in the TL regime can be investigated at $x=L>x_{\mathrm{%
H\uparrow }}$, and e-e scattering processes can be studied at $x=L<x_{%
\mathrm{H\uparrow }}$. The current $I_{\mathrm{det}}$ through the QD or QPC
is measured with a bias voltage $V_{\mathrm{D}}$ on the detector (D) ohmic
contact. The base current $I_{\mathrm{B}}$ is monitored to ensure $I_{%
\mathrm{inj}}\simeq I_{\mathrm{det}}+I_{\mathrm{B}}$ [see Fig 10].

\begin{figure}[tbp]
\begin{center}
\includegraphics[width = 3.3in]{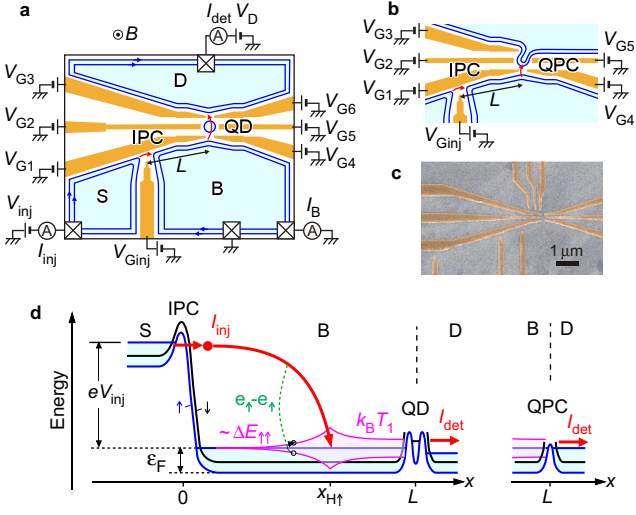}
\end{center}
\caption{\textbf{Schematic measurement setup.} \textbf{a} Device structure.
By applying gate voltages on the gates (yellow), an injector point contact
(IPC) and a detector quantum dot (QD) are formed between the source (S),
base (B), and drain (D) regions. The blue parallel lines show edge channels
at $\protect\nu =2$. \textbf{b} Device configuration for a quantum point
contact (QPC) detector. \textbf{c} Scanning electron micrograph with the
false color of a control device with several IPC gates. \textbf{d} Schematic
energy diagram of the measurement. Electrons are filled with the Fermi
energy $\protect\varepsilon _{\mathrm{F}}$ for up- and down-spin channels.
The energy-space trajectory of hot electrons is shown by the red arrows.
Electron-hole plasma near the chemical potential in B is analyzed by
measuring $I_{\mathrm{det}}$ through the QD or QPC.}
\end{figure}

\bigskip

\textbf{Energy-space trajectory of hot electrons.} First, we investigate the
hot-spot positions $x_{\mathrm{H\uparrow }}$ and $x_{\mathrm{H\downarrow }}$
by using QPC detection\cite{OtaPRB2019,AkiyamaAPL2019}. As shown in Fig. 3a,
the QPC conductance $G$ obtained without injecting hot electrons (by opening
the IPC with $V_{\mathrm{inj}}=$ 0 and $V_{\mathrm{Ginj}}=$ 0) shows
standard quantized conductances as a function of the gate voltage $V_{%
\mathrm{G4}}$. The onsets for spin-up and -down transport are seen at $V_{%
\mathrm{G4}}\sim -0.4$ V and $-0.2$ V, respectively. Additional peaks at $V_{%
\mathrm{G4}}\sim -0.5$ V are associated with parasitic impurity states,
which play minor roles in the following analysis. The following QPC
detection was performed at $V_{\mathrm{D}}=0$. When hot spin-up electrons at
energy $eV_{\mathrm{inj}}$ are injected from the IPC for $L=$ 1 $\mu $m,
distinct features appear in the detector current profile $I_{\mathrm{det}}$,
as shown in Fig. 3b. For the bottom trace at small $eV_{\mathrm{inj}}=$ 2
meV, a single current step with a height comparable to $I_{\mathrm{inj}}=$ 1
nA is seen at the opening of the spin-up transport ($V_{\mathrm{G4}}\sim
-0.4 $ V). This ensures no tunneling between spin-up and -down channels.
With increasing $eV_{\mathrm{inj}}$, current peaks (rather than steps) show
up for both spin-up and -down transports (in the pink stripes). As shown in
the insets (i) and (ii) to Fig. 3a, each peak can be understood by
considering the bolometric detection for each channel, where higher-energy
electrons are transmitted but lower-energy holes are reflected\cite%
{OtaPRB2019}. Therefore, the peak current is proportional to the density of
the electron-hole plasma in the respective channel, while the sensitivity
for up spins may be greater than that for down spins. Since the hot spot is
the point where the detected current is maximal, the strategy to determine
the hot spot position (for a given $L$) is to look for the voltage $V_{%
\mathrm{inj}}$ that maximizes the detected current. The peak currents for up
spins ($V_{\mathrm{G4}}=-0.42$ V) and down spins ($V_{\mathrm{G4}}=-0.18$ V)
are plotted as a function of $eV_{\mathrm{inj}}$ in Fig. 3c. The spin-up
current is maximized at $eV_{\mathrm{inj}}$ = 22 meV, which indicates that
the position of the hot spot in the spin-up channel coincides with that of
the QPC detector ($x_{\mathrm{H\uparrow }}=$ 1 $\mu $m for $eV_{\mathrm{inj}%
} $ = 22 meV). Similarly, the spin-down current is maximized at $eV_{\mathrm{%
inj}}$ = 14 meV, where the hot spot in the spin-down channel is considered
to be located at the position of the QPC detector ($x_{\mathrm{H\downarrow }%
}=$ 1 $\mu $m for $eV_{\mathrm{inj}}$ = 14 meV).

\begin{figure}[tbp]
\begin{center}
\includegraphics[width = 3.3in]{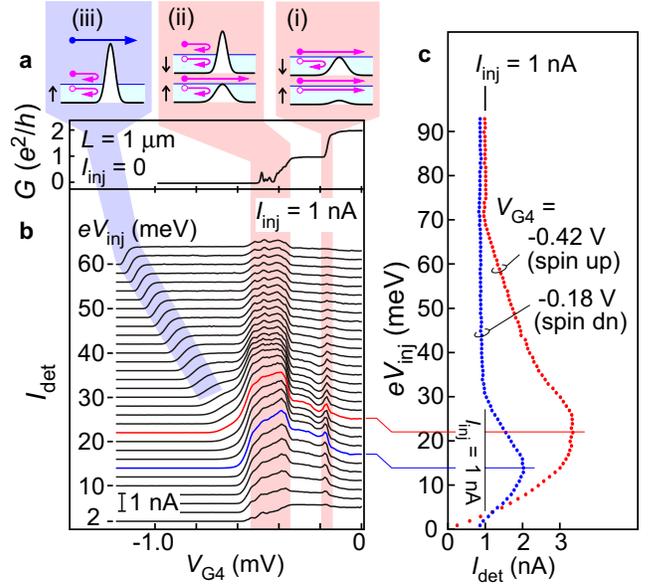}
\end{center}
\caption{\textbf{Quantum point contact (QPC) detection of electron-hole
plasma. } \textbf{a} Conductance $G=I_{\mathrm{det}}/V_{\mathrm{D}}$ of the
detector QPC measured with $V_{\mathrm{D}}=$ 0.1\ mV. Transmission and
reflection of hot electrons (the blue arrow) and cold electrons and holes
(the magenta arrows) are illustrated in the insets (i), (ii), and (iii). 
\textbf{b} $I_{\mathrm{det}}$ profiles obtained at various $eV_{\mathrm{inj}%
} $. $V_{\mathrm{G4}}$ in the horizontal axis was simultaneously swept with $%
V_{\mathrm{G5}}$ ($=V_{\mathrm{G4}}$). \textbf{c} The peak current of $I_{%
\mathrm{det}}$ for spin-up (red) and -down (blue) are plotted as a function
of $eV_{\mathrm{inj}}$. The maximum peak current appears when the hot spot
is located close to the detector.}
\end{figure}

We repeated similar measurements for various $L$, and the obtained hot-spot
positions are summarized in Fig. 4a. In addition to the data obtained with
the QPC (the circles), data obtained with the QD only for spin-up (described
below) are also plotted by the squares. Both data for up spins agree well
with each other within the experimental reproducibility. The smooth
connection of the data points can be used to extract the energy-space
trajectory of hot electrons. $\gamma _{\uparrow \uparrow }$ can be obtained
from $\gamma _{\uparrow \uparrow }=d\left( eV_{\mathrm{inj}}\right) /dL$, as
the hot spot $x_{\mathrm{H\uparrow }}$ ($=L$) increases with $eV_{\mathrm{inj%
}}$ at rate $\gamma _{\uparrow \uparrow }^{-1}$. Empirically, the power-law
dependence in a form of $\gamma _{\uparrow \uparrow }=aE_{\mathrm{h}%
}^{-\lambda }$ for hot-electron energy $E_{\mathrm{h}}$ explains the data
with parameters $a$ and $\lambda $\cite{OtaPRB2019}. This suggests an energy(%
$E_{\mathrm{h}}$)-space($x$) trajectory%
\begin{equation}
E_{\mathrm{h}}\left( x\right) =\left[ \left( eV_{\mathrm{inj}}\right)
^{\lambda +1}-a\left( \lambda +1\right) x\right] ^{1/\left( \lambda
+1\right) },
\end{equation}%
and the hot-spot condition%
\begin{equation}
eV_{\mathrm{inj}}=\left[ a\left( \lambda +1\right) x_{\mathrm{H\uparrow }}%
\right] ^{1/\left( \lambda +1\right) }.
\end{equation}%
The latter reproduces our data very well with $\lambda =1.8$, as shown by
the red curve in Fig. 4a. Corresponding energy loss rate $\gamma _{\uparrow
\uparrow }\left( E_{\mathrm{h}}\right) $ increases with decreasing $E_{%
\mathrm{h}}$, as shown in Fig. 4b. The energy-space trajectory $E_{\mathrm{h}%
}\left( x\right) $ is shown for several $eV_{\mathrm{inj}}$ values in Fig.
4c. As the energies of incident electrons should be distributed at around $%
eV_{\mathrm{inj}}$ mostly due to the energy-dependent tunneling probability
of the IPC, possible dispersion of the trajectories for $\pm $1 meV
distribution is shown by pink stripes. Notice that hot-electron relaxation
proceeds rapidly just before the hot spot $x_{\mathrm{H\uparrow }}$, which
can be tuned with $V_{\mathrm{inj}}$. This is useful for investigating the
decay length of non-thermal states, as one can change the distance $D$
between the hot spot and the detector.

\begin{figure}[tbp]
\begin{center}
\includegraphics[width = 3.3in]{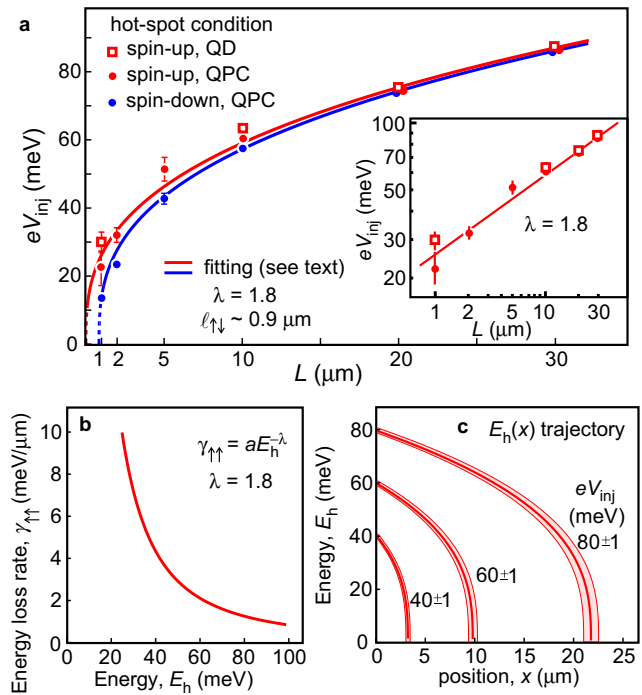}
\end{center}
\caption{\textbf{Energy-space trajectory of hot electrons.} \textbf{a} The
hot-spot conditions for up (red) and down (blue) spins measured with \textbf{%
\ }quantum point contact (QPC)\textbf{\ }(circles) and quantum dot (QD)%
\textbf{\ }(squares) detectors. The error bar shows the variations due to
the injector and detector conditions. The solid lines are fitted to the data
with $\protect\lambda =1.8$ (the uncertainty between 1.6 and 2.2) and $\ell
_{\uparrow \downarrow }=0.9$ $\protect\mu $m. The inset shows the data in
the double logarithmic plot. (b) The estimated energy loss rate $\protect%
\gamma _{\uparrow \uparrow }$ by assuming the power dependence of
hot-electron energy $E_{\mathrm{h}}$. (c) The estimated energy($E_{\mathrm{h}%
}$)-space($x$) trajectory of hot electrons for $eV_{\mathrm{inj}}=$ 40, 60,
and 80 meV. Each stripe represents the spread associated with variation of $%
eV_{\mathrm{inj}}$ by $\pm $1 meV.}
\end{figure}

As for the spin-down hot spot $x_{\mathrm{H\downarrow }}$ shown by the blue
circles in Fig. 4a, the $L$ dependence can be reproduced by the blue curve
which is obtained just by shifting the red curve horizontally by $\ell _{%
\mathrm{\uparrow \downarrow }}$ = 0.9 $\mu $m. This suggests that finite
length $\ell _{\mathrm{\uparrow \downarrow }}$, independent of $eV_{\mathrm{%
inj}}$, is required for exciting spin-down electrons through the
inter-channel interaction.\ This $\ell _{\mathrm{\uparrow \downarrow }}$ is
much longer than the length $l_{\mathrm{SC}}$ required for spin-charge
separation in the TL regime. Similar devices show $l_{\mathrm{SC}}\simeq $
0.18 $\mu $m at $eV_{\mathrm{inj}}=$ 0.6 meV, and $l_{\mathrm{SC}}$
decreases with increasing $eV_{\mathrm{inj}}$ ($l_{\mathrm{SC}}\propto 1/V_{%
\mathrm{inj}}$)\cite{Itoh-PRL2018}. The difference ($\ell _{\mathrm{\uparrow
\downarrow }}>l_{\mathrm{SC}}$) indicates the existence of the stochastic
inter-channel e-e scattering regime between $x_{\mathrm{H\uparrow }}$ and $%
x_{\mathrm{H\downarrow }}$.

The transition between intra- and inter-channel scattering regimes can be
confirmed from the data in Fig. 3c. In the range of $eV_{\mathrm{inj}}$
between 30 and 70 meV for $L<x_{\mathrm{H\uparrow }}$, spin-up electrons are
highly excited ($I_{\mathrm{det}}>I_{\mathrm{inj}}$) but no spin-down
electrons are excited ($I_{\mathrm{det}}\simeq I_{\mathrm{inj}}$) because
the intra-channel scattering is dominant. In contrast, both signals become
large at $eV_{\mathrm{inj}}<$ 25 meV and change in a similar manner at $eV_{%
\mathrm{inj}}<$ 15 meV ($L>x_{\mathrm{H\downarrow }} $), where the
inter-channel interaction equilibrates the two channels. The sequence of the
intra-channel scattering regime [from (i) to (iii) in Fig. 1a] followed by
the inter-channel scattering regime [(iii)-(iv)] is justified with the
result.

\bigskip

\textbf{QD spectroscopy. }We investigate electronic excitation in the
spin-up channel by using the QD detector\cite%
{Altimiras-NatPhys10,Itoh-PRL2018}. Coulomb diamond characteristics of the
QD are shown in Fig. 11. Figure 5a shows the Coulomb blockade (CB)
oscillations obtained with $V_{\mathrm{D}}=$ 0.2 mV under hot-electron
injection at $L$ = 10 $\mu $m. The horizontal axis $\Delta V_{\mathrm{G5}}$
is the relative voltage of $V_{\mathrm{G5}}$ from the left onset of the
right CB peak. The width of the CB peaks in the bottom trace measures the
transport window of $eV_{\mathrm{D}}$. When hot electrons are injected at
various $eV_{\mathrm{inj}}$ and fixed $I_{\mathrm{inj}}=$ 1.8 pA, excess
current appears in the entire CB region, as highlighted by the pink color.
This current measures the electrons excited above the QD level. For
instance, the current at $\Delta V_{\mathrm{G5}}=$ $-$19 mV (the dashed
line) becomes maximum at $eV_{\mathrm{inj}}=$ 62 meV, as shown in Fig. 5b.
This measures the spin-up hot-spot condition, where the hot electrons
injected at $eV_{\mathrm{inj}}=$ 62 meV just relaxed to the Fermi energy at $%
L$ = 10 $\mu $m. This coincides with the QPC data, as shown by the squares
in Fig. 4a. In the following analysis, we use the hot-electron trajectory
given by Eq. (2) at $\lambda =1.8$, while the parameter $a$ for each
geometry ($L$) is determined from $eV_{\mathrm{inj}}$ that maximizes $I_{%
\mathrm{det}}$ by using Eq. (3).

\begin{figure}[tbp]
\begin{center}
\includegraphics[width = 3.3in]{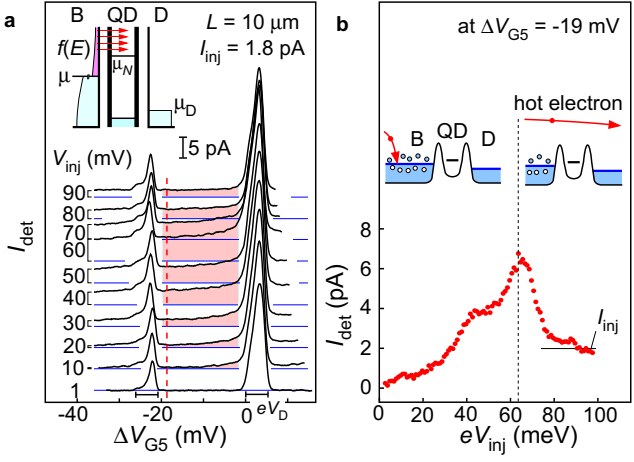}
\end{center}
\caption{\textbf{Quantum dot (QD) detection of electron-hole plasma. }%
\textbf{a} Coulomb oscillations of the QD observed in detector current $I_{%
\mathrm{det}}$ as a function of relative gate voltage $\Delta V_{\mathrm{G5}%
} $ under hot-electron injection at various injection voltage\textbf{\ }$V_{%
\mathrm{inj}}$. The excess current is highlighted by the pink color. The
inset shows the energy diagram of the QD. \textbf{b} $I_{\mathrm{det}}$
measured in the deep Coulomb blockade (CB)\textbf{\ }region at $\Delta V_{%
\mathrm{G5}}=$ -19 mV [the dashed line in \textbf{a}]. The left inset shows
the energy diagram for $eV_{\mathrm{inj}}\lesssim $ 60 meV, where the hot
electrons relax before reaching the QD. The right inset shows the diagram
for $eV_{\mathrm{inj}}\gtrsim $ 60 meV, where the hot electrons pass over
the QD.}
\end{figure}

The broad current profile in Fig. 5a suggests that the electrons are excited
well above the charging energy of the QD ($U\simeq $ 1.5 meV) as well as the
single-particle energy spacing of the QD ($\bar{\Delta}\simeq $ 0.2 meV on
average). The current decreases with decreasing $\Delta V_{\mathrm{G5}}$
reflecting the electron energy distribution function $f\left( E\right) $ of
the channel. To evaluate $f\left( E\right) $, the current profiles are
plotted in the logarithmic scale in Fig. 6a for $eV_{\mathrm{inj}}\leq $ 60
meV, where the hot electrons are fully relaxed before reaching the QD\
detector ($x_{\mathrm{H\uparrow }}<L$). The current profile in the CB region
shows exponential dependence $I_{\mathrm{det}}\propto \exp \left( \alpha
\Delta V_{\mathrm{G5}}/W_{\mathrm{c}}\right) $ as shown by the dashed lines.
Here, $\alpha \simeq 0.044e$ is the lever arm factor to convert $\Delta V_{%
\mathrm{G5}}$ into the electrochemical potential $\mu _{N}=-\alpha \Delta V_{%
\mathrm{G5}}$ of the $N$-electron QD, and $W_{\mathrm{c}}$ is the
characteristic energy of the current profile. For $eV_{\mathrm{inj}}>$ 62
meV, the hot electrons pass over the QD potential ($x_{\mathrm{H\uparrow }}>L
$), as shown in the inset to Fig. 6b. Noting that all the hot electrons are
eventually absorbed by the ammeter, we plot in Fig. 6b $I_{\mathrm{det}}-I_{%
\mathrm{inj}}$ as the net current in the logarithmic scale. The exponential
profile, $\propto \exp \left( \alpha \Delta V_{\mathrm{G5}}/W_{\mathrm{c}%
}^{\prime }\right) $, in the CB region is characterized by $W_{\mathrm{c}%
}^{\prime }$. We confirmed that $W_{\mathrm{c}}$ and $W_{\mathrm{c}}^{\prime
}$ do not change significantly with $I_{\mathrm{inj}}$, as shown in Figs. 6c
and 6d, as well as $N$ [partially seen at $\Delta V_{\mathrm{G5}}<-30$ mV in
Figs. 6a and 6b]. Some other data obtained at different $L$ and $I_{\mathrm{%
inj}}$ are shown in Fig. 12.

To analyze the exponential tail in the CB region, we rely on the orthodox CB
theory with a continuous density of QD states\cite{BookNazarov}, as all
features associated with the excited states of the QD are smeared out by the
broad profile. We also assume the QD is relaxed to the ground state before
accepting a single hot-electron. Under these assumptions, the detector
current yields 
\begin{equation}
I_{\mathrm{det}}=\int_{\mu _{N}}^{\infty }\frac{G_{\mathrm{QD}}}{e}f\left(
E\right) dE
\end{equation}%
with $G_{\mathrm{QD}}$ the tunneling conductance (see Methods). If $G_{%
\mathrm{QD}}\propto e^{E/W_{\mathrm{t}}}$ increases exponentially with
energy $E$ at characteristic energy $W_{\mathrm{t}}$ of the tunneling, the
model suggests exponential energy dependence of the distribution function $%
f\left( E\right) \propto \exp \left( -E/W_{\mathrm{d}}\right) $ with $W_{%
\mathrm{d}}^{-1}=$ $W_{\mathrm{t}}^{-1}+W_{\mathrm{c}}^{-1}$. While we do
not know $W_{\mathrm{t}}$ of our device, the spread of the current tail ($W_{%
\mathrm{c}}$) should reflect the spread of the distribution function ($W_{%
\mathrm{d}}$) when $W_{\mathrm{c}}$ is not too large.

\begin{figure}[tbp]
\begin{center}
\includegraphics[width = 3.3in]{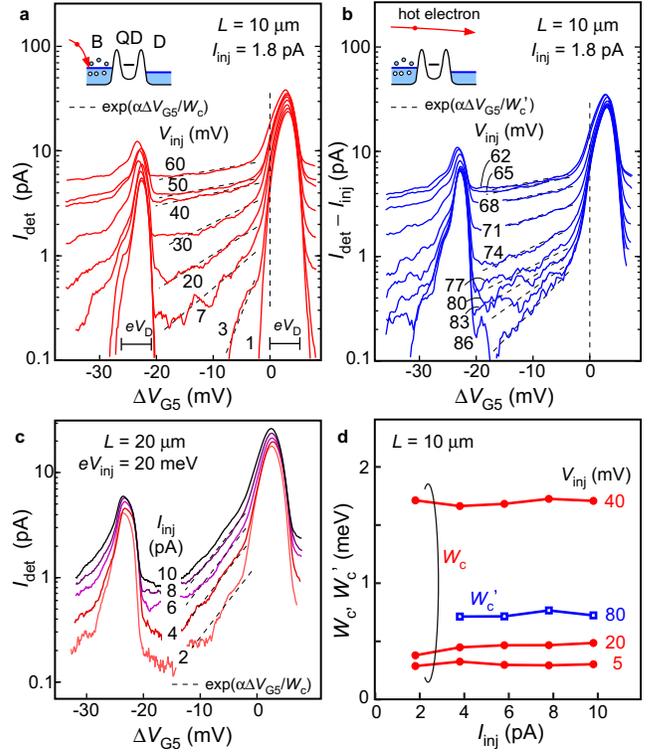}
\end{center}
\caption{\textbf{Quantum dot (QD) spectroscopy for energy distribution
functions.} \textbf{a} Detector current $I_{\mathrm{det}}$ as a function of
relative gate voltage $\Delta V_{\mathrm{G5}}$ at various injection energy%
\textbf{\ }$eV_{\mathrm{inj}}\leq $ 60 meV for studying non-thermal
metastable states after long transport. \textbf{b} Excess current \textbf{\ }
$I_{\mathrm{det}}-I_{\mathrm{inj}}$ for $eV_{\mathrm{inj}}\geq $ 62 meV for
studying non-equilibrium states in the course of electron-electron
scattering. The dashed lines, $\propto \exp \left( \protect\alpha V_{\mathrm{%
G5}}/W_{\mathrm{c}}\right) $ in \textbf{a} and $\propto \exp \left( \protect%
\alpha V_{\mathrm{G5}}/W_{\mathrm{c}}^{\prime }\right) $ in \textbf{b}, are
fitted to the data. The current tails in \textbf{a} and \textbf{b} are
characterized by cut-off energies $W_{\mathrm{c}}$ and $W_{\mathrm{c}%
}^{\prime }$, respectively. The insets show the energy diagrams around the
QD. \textbf{c} $I_{\mathrm{det}}$ for several injection current $I_{\mathrm{%
inj}}$ at $eV_{\mathrm{inj}}=$ 20 meV. No significant change in the slope
(the dashed lines) is seen. \textbf{d} Injection current $I_{\mathrm{inj}}$
dependence of \textbf{\ }cut-off energies $W_{\mathrm{c}}$ (red) and $W_{%
\mathrm{c}}^{\prime }$ (blue). }
\end{figure}

\bigskip

\textbf{Non-thermal metastable state.} The state after equilibration at $%
x>x_{\mathrm{H\downarrow }}$ can be studied with the data in Fig. 6a. The
coexistence of the small exponential current tail with large $W_{\mathrm{c}}$
and the large CB current peak with a sharp onset associated with cold
electrons manifests the non-thermal metastable state of the TL liquid\cite%
{WashioPRB2016,Itoh-PRL2018,RodriguezNatComm2020}. In the present case, the
non-thermal states are analyzed with the binary form of Eq. (1). The low
temperature ($T_{0}$) is always close to the base temperature ($T_{\mathrm{B}%
}$), as the CB peak remains sharp at any $V_{\mathrm{inj}}$ in Fig. 6a. The
high temperature ($T_{1}$) is related to the spread of the current tail ($k_{%
\mathrm{B}}T_{1}\sim W_{\mathrm{c}}$). Significantly different temperatures (%
$T_{1}\gg T_{0}$) are seen at higher $V_{\mathrm{inj}}$. These current
profiles are qualitatively the same as those reported for small excitation
energies (10 - 100 $\mu $eV) where the low-energy TL physics governs.
Surprisingly, low-energy physics emerges from the high-energy source (1 - 60
meV). Note that the non-thermal metastable state is distinct from the
ordinary non-equilibrium state often seen in higher dimensions\cite%
{BookAltshulerAronovEE,PothierThermalize}.

As several QD levels are involved in the transport for the highly
non-equilibrium state, $p$ in Eq. (1) cannot be estimated from the current
profile. Instead, $p$ can be obtained from the electrical power $P_{\mathrm{%
inj}}=I_{\mathrm{inj}}V_{\mathrm{inj}}$ injected from the IPC. It is known
that a single edge channel carries heat current $J=\frac{\pi ^{2}}{6h}\left(
k_{\mathrm{B}}T\right) ^{2}$ if the electrons are in thermal equilibrium at
temperature $T$\cite{JezouinScience2013}. Therefore, if the spin-up and
-down channels were thermally equilibrated with $P_{\mathrm{inj}}$, the
thermal energy would increase to $k_{\mathrm{B}}T_{\mathrm{th}}=\sqrt{\frac{%
3h}{\pi ^{2}}P_{\mathrm{inj}}}$ by neglecting the thermal energy of the base
temperature ($T_{\mathrm{B}}$). This yields $k_{\mathrm{B}}T_{\mathrm{th}}=$
10 $\mu $eV for $P_{\mathrm{inj}}=$ 13 fW ($I_{\mathrm{inj}}=$ 1.8 pA and $%
V_{\mathrm{inj}}=$ 7 mV), which is much smaller than $W_{\mathrm{c}}=$ 0.3
meV at $V_{\mathrm{inj}}=$ 7 mV for the third trace from the bottom in Fig.
6a. The heat conservation suggests that only a small fraction, about $%
p=\left( k_{\mathrm{B}}T_{\mathrm{th}}/W_{\mathrm{c}}\right) ^{2}\simeq $ 10$%
^{-3}$ of electrons, is highly excited to the high-temperature $T_{1}$. Such
an unusual energy distribution is a consequence of the crossover from hot
electrons.

The exponential current tail in Fig. 6a starts to develop from $eV_{\mathrm{%
inj}}\simeq $ 2 meV, and the corresponding $W_{\mathrm{c}}$ value rapidly
increases with $eV_{\mathrm{inj}}$. The $W_{\mathrm{c}}$ values obtained at
various $eV_{\mathrm{inj}}$ and $L$ are summarized in Fig. 7. First, we
focus on the low $eV_{\mathrm{inj}}$ ($\lesssim $ 8 meV) regime for the
shortest $L=$ 1 $\mu $m, where $W_{\mathrm{c}}$ increases linearly with $eV_{%
\mathrm{inj}}$ (the dashed line). As the hot electron relaxes within a short
propagation length ($x_{\mathrm{H\uparrow }}<0.1$ $\mu $m as shown in the
auxiliary scale) in this case, the distribution function should measure the
non-thermal metastable state after traveling of $\sim $ 1 $\mu $m. Actually,
the slope $W_{\mathrm{c}}/eV_{\mathrm{inj}}=0.2$ is comparable to the values
obtained in the low-energy limit of the TL regime ($W_{\mathrm{c}}/eV_{%
\mathrm{inj}}=0.17$ for $W_{\mathrm{c}}=$ 34 $\mu $eV at $eV_{\mathrm{inj}}=$
200 $\mu $eV with QPC injection in Ref. \cite{Itoh-PRL2018}, $W_{\mathrm{c}%
}/eV_{\mathrm{inj}}=0.22-0.25$ for $W_{\mathrm{c}}=$ 14 - 17 $\mu $eV at $%
eV_{\mathrm{inj}}=$ 63 $\mu $eV with QD injection in Ref. \cite%
{RodriguezNatComm2020}, and $W_{\mathrm{c}}/eV_{\mathrm{inj}}=0.17$ for $W_{%
\mathrm{c}}=$ 20 - 35 $\mu $eV at $eV_{\mathrm{inj}}=$ 100 - 200 $\mu $eV
with indirect QPC injection in Ref. \cite{WashioPRB2016}). Our experiment
confirms that the same ratio ($W_{\mathrm{c}}/eV_{\mathrm{inj}}$) holds up
to $W_{\mathrm{c}}\simeq $ 1 meV, which is more than 10 times greater than
the values in the previous works. The linearity implies that the injection
energy fits in the energy range of the TL liquid without experiencing e-e
scattering. All electrons within the range interact with each other to form
the collective plasmon mode, as discussed in Fig. 1e. The memory of the
initial state at $eV_{\mathrm{inj}}$ remains in the broad electronic
distribution function with $W_{\mathrm{d}}$ ($\simeq W_{\mathrm{c}}\simeq
0.2eV_{\mathrm{inj}}$). This can be understood with the integrable nature of
holding many conserved quantities.

\begin{figure}[tbp]
\begin{center}
\includegraphics[width = 3in]{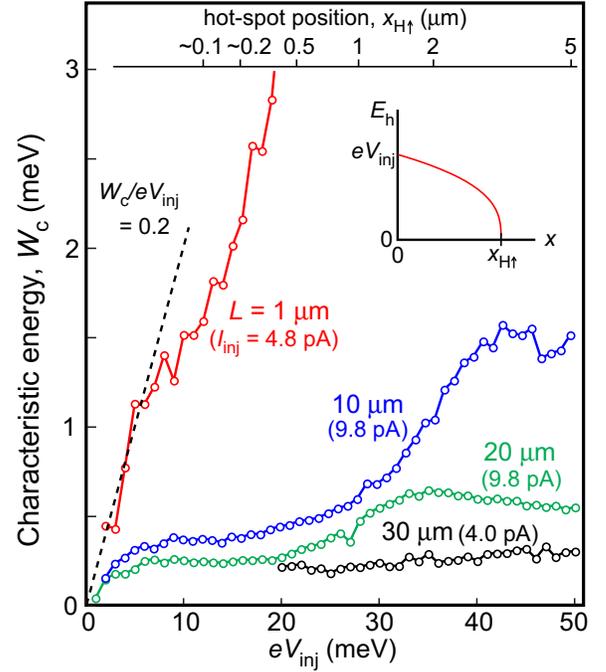}
\end{center}
\caption{\textbf{The cutoff energy }$W_{\mathrm{c}}$\textbf{\ for the
non-thermal states.} The dashed line shows the \textbf{\ }linear relation, $%
W_{\mathrm{c}}\simeq 0.2eV_{\mathrm{inj}}$, obtained in the low-energy limit
of Tomonaga-Luttinger (TL) liquids. The inset shows the energy-space
trajectory of hot electrons. The hot-spot position, $x_{\mathrm{H\uparrow }}$%
, obtained with Eq. (3) at $\protect\lambda =$ 1.8 is shown in the upper
scale. }
\end{figure}

However, $W_{\mathrm{c}}$ deviates from the linear dependence to lower
values at $eV_{\mathrm{inj}}\gtrsim 8$ meV for $L=$ 1 $\mu $m, as seen in
Fig. 7. $W_{\mathrm{c}}$ obtained with longer $L$ may follow the linear
dependence at small $eV_{\mathrm{inj}}\lesssim 1$ meV and becomes
substantially smaller at larger $eV_{\mathrm{inj}}$. The reduction in $W_{%
\mathrm{c}}$ might originate from the dissipation\cite%
{leSueurPRL2010,ProkudinaPRL2014} or the departure from the TL regime. To
characterize the decay of non-thermal states, the $W_{\mathrm{c}}$ data are
plotted as a function of the distance $D=L-x_{\mathrm{H\uparrow }}$ from the
hot spot to the detector, as shown in Fig. 8. The corresponding $eV_{\mathrm{%
inj}}$ values are shown in the auxiliary scale for each $L$. Each trace
features a gradual reduction of $W_{\mathrm{c}}$ (except for $L= $ 1 $\mu $%
m) at $eV_{\mathrm{inj}}>$ 50 meV followed by a sudden reduction at $eV_{%
\mathrm{inj}}<$ 10 meV ($D\simeq L$). Both features can be explained
consistently with the energy-dependent dissipation during traveling.

For $eV_{\mathrm{inj}}<$ 10 meV, the electron-hole excitation at the hot
spot is developing with increasing $eV_{\mathrm{inj}}$, while $D$ ($\simeq L$%
) is almost constant for each $L$. Therefore, the dissipation can be studied
by varying $L$ for a given $eV_{\mathrm{inj}}$, as shown by the two dotted
lines connecting the data points for $eV_{\mathrm{inj}}$ = 5 and 10 meV. The
steeper slope at higher $W_{\mathrm{c}}$ suggests increased dissipation at
higher excitation energy.

In contrast, the excitation at the hot spot does not increase further for $%
eV_{\mathrm{inj}}>$ 30 meV. As seen in the energy-space trajectory of Fig.
4c, the excess energy beyond 20 - 30 meV is lost during the long-distance
transport with the remaining 20 - 30 meV used to build the hot spot with the
highest excitation energy. Therefore, the $eV_{\mathrm{inj}}$ dependence at $%
eV_{\mathrm{inj}}>$ 30 meV in Fig. 8 should be understood as representing
the $D$ dependence of $W_{\mathrm{c}}$ . The overall feature at $eV_{\mathrm{%
inj}}>$ 50 meV follows a monotonic reduction of $W_{\mathrm{c}}$ with $D$,
as shown by the long dashed line. This slope is comparable to the slope of
the short-dashed lines if the data in the same range of $W_{\mathrm{c}}$ are
compared. The data shows how the electronic distribution function changes
with transport. The characteristic energy $W_{\mathrm{d}}$ ($\simeq W_{%
\mathrm{c}}$) decays fast if $W_{\mathrm{d}}$ is large (the decay length $%
\ell _{\mathrm{d}}\sim 5$ $\mu $m at $W_{\mathrm{c}}\simeq $ 1 meV), but the
decay length becomes longer for smaller $W_{\mathrm{d}}$ ($\sim $ 30 $\mu $m
at 0.2 meV). This is comparable to $\ell _{\mathrm{d}}\sim $ 20 $\mu $m
obtained from the low-energy experiment at $W_{\mathrm{c}}=$ 34 $\mu $eV\cite%
{Itoh-PRL2018}. The energy dependent $\ell _{\mathrm{d}}$ may be explained
by energy-dependent dissipation associated with the characteristics of the
environment. If the reduction signifies the departure of the TL regime, the
energy range for the TL physics could be $\Delta E_{\mathrm{TL}}\sim $ 1 meV
in our device.

\begin{figure}[tbp]
\begin{center}
\includegraphics[width = 3in]{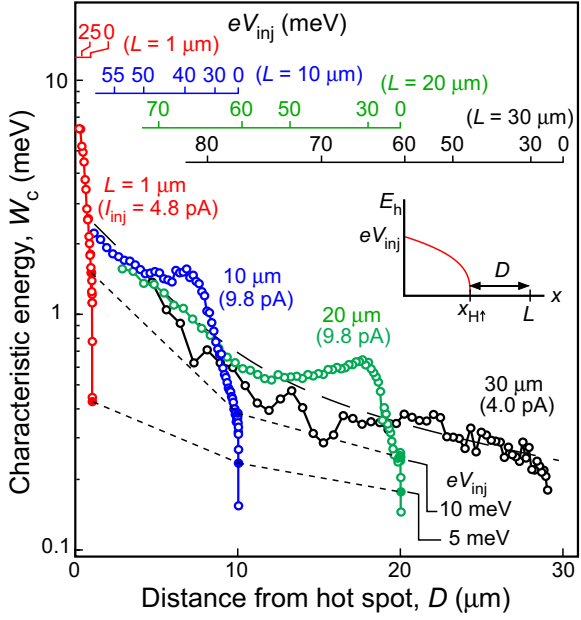}
\end{center}
\caption{\textbf{Decay of non-thermal metastable states. } The
characteristic energy\textbf{\ }$W_{\mathrm{c}}$\textbf{\ }is plotted as a
function of the distance $D$\ from the hot spot to the quantum dot (QD). $D$
is obtained from Eq. (3) at $\protect\lambda =$ 1.8. The inset shows the
energy-space trajectory. The $eV_{\mathrm{inj}}$ value for each $L$ is shown
in the respective auxiliary scale. The decay of $W_{\mathrm{c}}$ is shown by
the short dashed lines connecting the data points obtained with the same $%
eV_{\mathrm{inj}}$ ($\leq $ 10 meV) and the long dashed line as a guide to
the eye for the data at $eV_{\mathrm{inj}}\gtrsim $ 50 meV. }
\end{figure}

\bigskip

\textbf{e-e scattering. }The electron-hole plasma generated in the course of
the e-e scattering was analyzed with the current profile, $I_{\mathrm{det}%
}-I_{\mathrm{inj}}$, at $x_{\mathrm{H\uparrow }}>L$, in Fig. 6b. The data
indicates that $W_{\mathrm{c}}^{\prime }$ (the inverse of the slope)
decreases with increasing $eV_{\mathrm{inj}}$. Considering that the energy
loss rate $\gamma _{\uparrow \uparrow }$ increases as $E_{\mathrm{h}}$
decreases during the transport from the IPC to the detector [see Fig. 4b],
it is likely that the obtained $W_{\mathrm{c}}^{\prime }$ value reflects
most sensitively the e-e scattering near the detector QD. From this point of
view, we plot in Fig. 9 $W_{\mathrm{c}}^{\prime }$ as a function of $E_{%
\mathrm{h}}\left( L\right) $, the hot-electron energy at the detector
position. The two data sets for $L=$ 1 and 10 $\mu $m show similar
dependency, which justifies this approach. The data show that $W_{\mathrm{c}%
}^{\prime }$ decreases with increasing $E_{\mathrm{h}} $, as shown by the
dashed line. Namely, the electron-hole plasma is highly excited in the
vicinity of the hot spot, as illustrated in the inset.

\begin{figure}[tbp]
\begin{center}
\includegraphics[width = 3in]{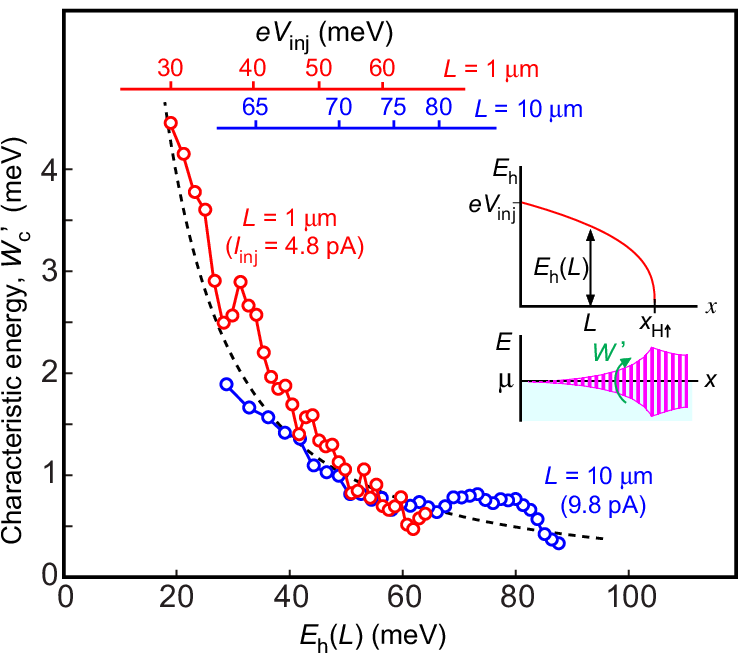}
\end{center}
\caption{\textbf{The characteristic energy }$W_{\mathrm{c}}^{\prime }$ 
\textbf{for the electron-electron scattering.\ }The hot-electron energy%
\textbf{\ }$E_{\mathrm{h}}\left( L\right) $ at the quantum dot (QD) detector
is obtained from Eq. (2) at $\protect\lambda =$ 1.8. The $eV_{\mathrm{inj}}$
value for each $L$ is shown in the respective auxiliary scale. The dashed
line is a guide to the eye by using a function proportional to $E_{\mathrm{h}%
}^{-1.5}$. The inset shows the energy-space trajectory and buildup of
electron-hole plasma.}
\end{figure}

The physical meaning of $W_{\mathrm{c}}^{\prime }$ can be understood by
considering the electron-hole generation process. Suppose that an
intra-channel e-e scattering between a hot electron with energy $E_{\mathrm{h%
}}$ and a cold electron with $E_{\mathrm{c}}$ ($<0$) generates an
electron-hole pair with a hole at $E_{\mathrm{c}}$ and an electron near the
Fermi energy [at $E_{\mathrm{c}}+\varepsilon $ ($>0$)] by the energy
exchange $\varepsilon $, as shown in Fig. 1c. Scattering with various
combinations of $E_{\mathrm{c}}$ and $\varepsilon $ induces the plasma.
Here, $\varepsilon $ is bounded by the momentum conservation. For a given $%
E_{\mathrm{h}}$, we assume that the scattering probability $P\left( E_{%
\mathrm{c}},\varepsilon ;E_{\mathrm{h}}\right) \varpropto e^{-\varepsilon
/\Delta E_{\uparrow \uparrow }}$ for $E_{\mathrm{c}}$ and $\varepsilon $
depends only on $\varepsilon $ with the upper bound $\Delta E_{\uparrow
\uparrow }$ that depends on $E_{\mathrm{h}}$. For the present case, we are
interested in the electronic energy distribution function $f\left( E\right) $
above the Fermi energy ($E>0$). We further assume that $f\left( E\right) $
is determined by the e-e scattering in the vicinity of the detector by
neglecting the scattering that occurs in the upstream region. Then, $f\left(
E\right) $ is given by the instantaneous e-e scattering with $E_{\mathrm{h}}$
at $x=L$ as

\begin{equation}
f\left( E\right) \propto \int_{-\infty }^{0}P\left( E_{\mathrm{c}%
},\varepsilon ;E_{\mathrm{h}}\right) dE_{\mathrm{c}}\propto e^{-E/\Delta
E_{\uparrow \uparrow }},  \label{fE}
\end{equation}%
where $E=E_{\mathrm{c}}+\varepsilon $ was used in the integration.
Therefore, provided that $W_{\mathrm{c}}^{\prime }$ measures the spread $W_{%
\mathrm{d}}$ of the distribution function, the obtained $W_{\mathrm{c}%
}^{\prime }$ can be understood as the energy cutoff $\Delta E_{\uparrow
\uparrow }$ of the intra-channel e-e scattering.

The monotonic reduction of $W_{\mathrm{c}}^{\prime }$ with $E_{\mathrm{h}%
}(L) $ in Fig. 9 is consistent with the concave dispersion in Fig. 1b, as
explained above. In this way, the energy loss of hot electrons ($E_{\mathrm{h%
}}>$ 20 meV) and excitation of cold electrons ($E_{\mathrm{h}}<$ 5 meV)\ are
consistently explained with the intra-channel e-e scattering. It should be
noted that the energy dependency (larger $W_{\mathrm{c}}^{\prime }$ for
smaller $E_{\mathrm{h}}$) of e-e scattering in Fig. 9 is opposite to that
(larger $W_{\mathrm{c}}$ for larger $eV_{\mathrm{inj}}$) for TL states in
Fig. 7.

\bigskip

\textbf{Conclusions}

We have investigated how high-energy hot electrons lose their energy by
exciting cold electrons. A QPC was used as a broad-band spin-dependent
bolometer to extract the energy-space trajectory of hot electrons. A QD was
employed as a narrow-band spectrometer to detect non-equilibrium electronic
excitation near the Fermi energy. The hot electrons lose their energy with
the e-e scattering and eventually create spin-dependent hot spots. The cold
electrons are maximally excited at the hot spot and equilibrate while
traveling the channels. Interestingly, the system always ends up with a
non-thermal metastable state without fully relaxing into thermal
equilibrium, even when the initial hot-electron energy is well beyond the TL
regime. This demonstrates the robustness of non-thermal states against the
excitation energy. The non-thermal state depends on the excitation energy
even after traveling 20 $\mu $m. This suggests that the final state
remembers the initial state even after the equilibration, in consistency
with the integrable nature with many conserved quantities. In addition, the
non-equilibrium state generated with the e-e scattering is in-situ
investigated to evaluate the maximum energy exchange in the scattering. The
analysis implies the significance of inter-channel scattering with different
spins as well as intra-channel scattering with the same spins. The
experiment provides a novel crossover from single particles to a
non-equilibrium many-body state. The scheme can be used for studying
non-linear TL liquids in the presence of non-linear dispersion\cite%
{ImambekovRMP2012}. The robustness of non-thermal states is attractive for
transporting information and energy for long distances\cite%
{CalzonaPRB2017,KaminishiPRA2018,UedaNRP2020}.

\bigskip

\textbf{Methods}

\textbf{Hot-electron injection.} We use custom-made voltage source and
ammeter to inject hot electrons. A large input impedance of 100 k$\Omega $
in the ammeter prevents unwanted large current during adjustments. $\ V_{%
\mathrm{Ginj}}$ is precisely adjusted by a software to obtain the desired $%
I_{\mathrm{inj}}$ value (1 pA - 1 nA) within a few \% error before measuring 
$I_{\mathrm{det}}$ at each condition ($V_{\mathrm{inj}}$, $V_{\mathrm{G4}}$,
etc.). $V_{\mathrm{Ginj}}$ depends strongly on $V_{\mathrm{inj}}$
(typically, from $V_{\mathrm{Ginj}}\simeq $ -1.1 V at $V_{\mathrm{inj}}=$ 5
mV to $V_{\mathrm{Ginj}}\simeq $ -1.9 V at $V_{\mathrm{inj}}=$ 80 mV with an
almost linear relation in between) for $I_{\mathrm{inj}}=$ 1 nA. Slightly
more negative $V_{\mathrm{Ginj}}$ by $\simeq $ -0.01 V was needed for $I_{%
\mathrm{inj}}=$ 10 pA. While the location of the injector should be pushed
toward the other gate G1 with more negative $V_{\mathrm{Ginj}}$, we
neglected possible change in transport length $L$. $I_{\mathrm{inj}}$ is
kept sufficiently low, 1 $-$ 10 pA for the QD detection to minimize the
heating inside the QD and $\sim $ 1 nA for the QPC detection for better
visibility.

As the e-e scattering is sensitive to the distance from the channel to the
gate metal, we fixed the gate voltages defining the channel ($V_{\mathrm{G1}%
}=V_{\mathrm{G3}}=$ -0.6 V and $V_{\mathrm{G2}}=$ -0.45 V) throughout this
paper. These gate voltages were chosen in such a way that the edge potential
becomes gentle enough to suppress unwanted LO phonon emission while high
enough to confine hot electrons\cite{OtaPRB2019,EmaryPRB2019,AkiyamaAPL2019}%
. While $V_{\mathrm{inj}}$ (1 - 100 meV) is much larger than the cyclotron
energy of about 9 meV at 5 T, we did not observe any tunneling transition to
higher Landau levels in the present conditions.

\textbf{QPC detection.} We focused on the current peaks appearing at the
onsets of the conductance steps of the QPC, as described in the main text.
Additional data that support our observation are shown in Supplementary Note
SN1. However, one can investigate the hot-electron energy from the current
step appearing at $V_{\mathrm{G4}}<-0.8$ V and $eV_{\mathrm{inj}}>$ 30 meV
in Fig. 3b. This current step appears when the potential barrier height
coincides with the hot-electron energy, as illustrated in the inset (iii) to
Fig. 3a, and disappears at $eV_{\mathrm{inj}}\lesssim $ 25 meV in
consistency with the hot spot condition obtained above. In our previous
report, the hot-electron energy is obtained from $V_{\mathrm{G4}}$ at the
current step by using a conversion factor determined from the LO phonon
replicas of the current step\cite{OtaPRB2019,AkiyamaAPL2019}. The obtained $%
\lambda $ in the range between 1 and 2 is consistent with the present value (%
$\lambda =1.8$), while $\lambda $ may depend on the device and the side gate
voltage. However, the hot-electron energy cannot be obtained for the present
case, as we suppressed LO phonon emission. Instead, the whole hot-electron
energy is used for generating the electron-hole plasma, which is used to
estimate $p$ in this study.

\textbf{QD detection.} The QD characteristics ($U\simeq $ 1.5 meV, $\bar{%
\Delta}\simeq $ 0.2 meV, and $\alpha \simeq 0.044e$) were obtained from
standard Coulomb diamond measurements at $V_{\mathrm{inj}}=0$ and $V_{%
\mathrm{Ginj}}=0$. The base electron temperature $T_{\mathrm{B}}\simeq $ 110
mK was also obtained from the onset of the CB peak. The backaction from the
change in $V_{\mathrm{Ginj}}$ and $V_{\mathrm{inj}}$ to the detector
conditions is present. While this is negligible for the QPC detection, the
backaction to the QD is significant, as the CB peaks were shifted by $\sim $
30 mV in $V_{\mathrm{G5}}$ for $L=$ 1 $\mu $m and $\sim $ 8 mV for $L=$ 10 $%
\mu $m when $V_{\mathrm{inj}}$ was changed from 1 to 90 mV under the
adjustment of $V_{\mathrm{Ginj}}$ for $I_{\mathrm{inj}}=$ 5 pA. The relative
value $\Delta V_{\mathrm{G5}}$ from the CB peak is used in the main text.
Additional data that support our observation are shown in Supplementary Note
SN2.

\textbf{Analysis of the current profiles.} We use the orthodox CB theory to
analyze the broad current profile in the CB region at $\mu _{N}>\mu >\mu _{%
\mathrm{D}}$ (see inset to Fig. 5a). As the current is sufficiently low, the
dot should remain close to the base temperature. In this case, the tunneling
rate from the edge channel of interest to the dot is given by $\Gamma _{%
\mathrm{dc}}\simeq \int_{\mu _{N}}^{\infty }\frac{1}{e^{2}}G_{\mathrm{L}%
}f\left( E\right) dE$ for the distribution function $f\left( E\right) $,
where the tunneling conductance $G_{\mathrm{L}}$ may depends on energy. By
assuming $T_{\mathrm{B}}=0$ for simplicity, the tunneling rates from the dot
to the channel and drain read $\Gamma _{\mathrm{cd}}\simeq \int_{\mu }^{\mu
_{N}}\frac{1}{e^{2}}G_{\mathrm{L}}dE$ and $\Gamma _{\mathrm{Dd}}\simeq
\int_{\mu _{\mathrm{D}}}^{\mu _{N}}\frac{1}{e^{2}}G_{\mathrm{R}}dE$, where $%
G_{\mathrm{L}}$ and $G_{\mathrm{R}}$ are tunneling conductances of the
respective barriers. By solving the master equation, we obtain the steady
current $I=e\Gamma _{\mathrm{dc}}\Gamma _{\mathrm{Dd}}/\left( \Gamma _{%
\mathrm{dc}}+\Gamma _{\mathrm{Dd}}+\Gamma _{\mathrm{cd}}\right) $. By
assuming energy-dependent conductances $G_{\mathrm{L}}\propto e^{E/W_{%
\mathrm{t}}}$ and $G_{\mathrm{R}}\propto e^{E/W_{\mathrm{t}}}$ with the same
characteristic energy $W_{\mathrm{t}}$ for the tunneling, we obtain $I$ $%
\simeq \int_{\mu _{N}}^{\infty }\frac{1}{e}G_{\mathrm{QD}}f\left( E\right)
dE $ with $G_{\mathrm{QD}}\propto e^{E/W_{\mathrm{t}}}$ by neglecting
factors that weakly depend on energy. This form was used in the analysis of
distribution functions.

\bigskip 

\textbf{Supplementary Notes }

SN\textbf{1: QPC detection}

We prepare the channel configuration by applying appropriate gate voltages,
as shown in Fig. 10a. $V_{\mathrm{G1}}=V_{\mathrm{G3}}=$ -0.6 V and $V_{%
\mathrm{G2}}=$ -0.45 V are fixed throughout this study. We inject hot
electrons from the source (S) through the injection point contact (IPC) to
the base (B) by applying voltage $-V_{\mathrm{inj}}$ (= 1 - 100 mV) to the
ohmic contact. The injection current $I_{\mathrm{inj}}$ (= 1 nA) is adjusted
by tuning $V_{\mathrm{Ginj}}$. The resulting excitation in the edge channel
of B is analyzed with the quantum point contact (QPC) located at a distance $%
L$ from the IPC. The conductance $G$ of the QPC is varied in the range from $%
G=$ 0 to $2e^{2}/h$ by changing $V_{\mathrm{G4}}$ and $V_{\mathrm{G5}}$
simultaneously ($V_{\mathrm{G4}}=V_{\mathrm{G5}}$). For each $V_{\mathrm{G4}}
$ ($=V_{\mathrm{G5}}$) setting, $V_{\mathrm{Ginj}}$ is readjusted to obtain
the desired $I_{\mathrm{inj}}$ by using a software feed-back control, and
then $I_{\mathrm{inj}}$, $I_{\mathrm{det}}$, and $I_{\mathrm{B}}$ are
recorded at nominally $V_{\mathrm{D}}=0$. Figure 10b-d show representative
results ($I_{\mathrm{inj}}$, $I_{\mathrm{det}}$, and $I_{\mathrm{B}}$)
obtained for various $V_{\mathrm{G4}}$ values at $V_{\mathrm{inj}}$ = 40 mV
and $L$ = 1 $\mu $m. One can confirm the relation $I_{\mathrm{inj}}$ $\simeq 
$ $I_{\mathrm{det}}+I_{\mathrm{B}}$ from the data. The remainder ($I_{%
\mathrm{inj}}$ $-$ $I_{\mathrm{det}}-I_{\mathrm{B}}$) should be drained to
the other ohmic contact to B, as the input impedance of the ammeter for $I_{%
\mathrm{B}}$ is finite ($\sim $ 1 k$\Omega $).

\begin{figure}[tbp]
\begin{center}
\includegraphics[width = 3.3in]{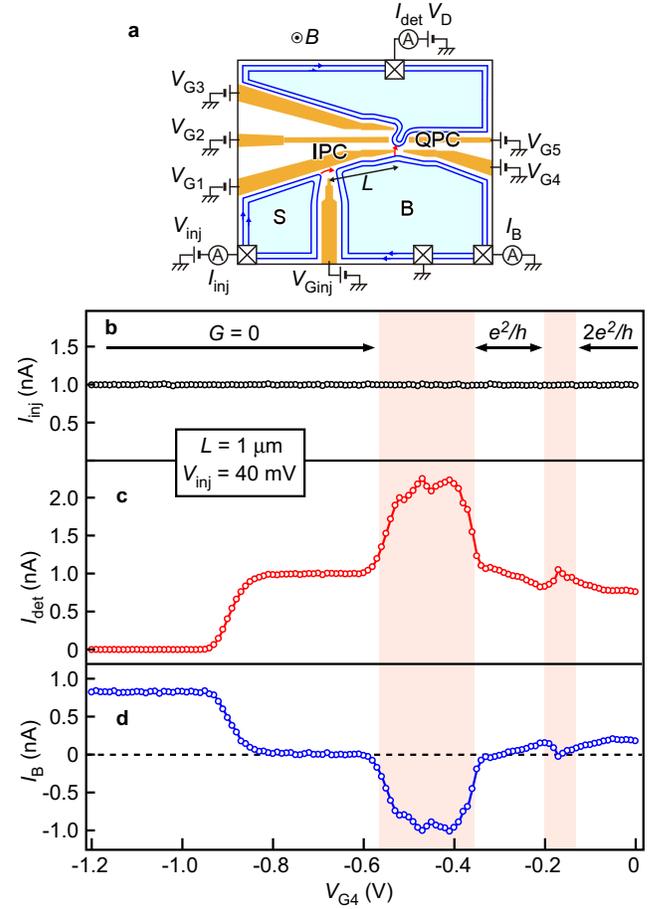}
\end{center}
\caption{\textbf{a} Schematic setup for the QPC detection under hot electron
injection from IPC. \textbf{b} $I_{\mathrm{inj}}$, \textbf{c} $I_{\mathrm{det%
}}$, and \textbf{d} $I_{\mathrm{B}}$ as a function of $V_{\mathrm{G4}}$. }
\end{figure}

The current step at $V_{\mathrm{G4}}\simeq $ -0.9 V measures the energy of
hot electrons. The hot electrons are fully reflected at $V_{\mathrm{G4}}<$
-0.9 V and fully transmitted through the QPC at $V_{\mathrm{G4}}>$ -0.9 V.
As the step height is identical to $I_{\mathrm{inj}}$, all hot electrons
injected from IPC are not fully relaxed to the Fermi energy. The current
peak at $V_{\mathrm{G4}}\simeq $ -0.5 V measures the electron-hole plasma in
the spin-up channel, which are generated by the intra-channel
electron-electron scattering. The other current peak at $V_{\mathrm{G4}%
}\simeq $ -0.2 V measures the electron-hole plasma in the spin-down channel,
which are generated by the inter-channel electron-electron scattering. These
scattering processes are discussed in the main paper. At $V_{\mathrm{G4}}=$
0 V, $I_{\mathrm{det}}$ is slightly smaller than $I_{\mathrm{inj}}$, and $I_{%
\mathrm{B}}$ becomes negative. This is an experimental artifact, in which
small but finite $V_{\mathrm{D}}$ (about 1 $\mu $V) was present in our
circuit.

\begin{figure}[tbp]
\begin{center}
\includegraphics[width = 3.3in]{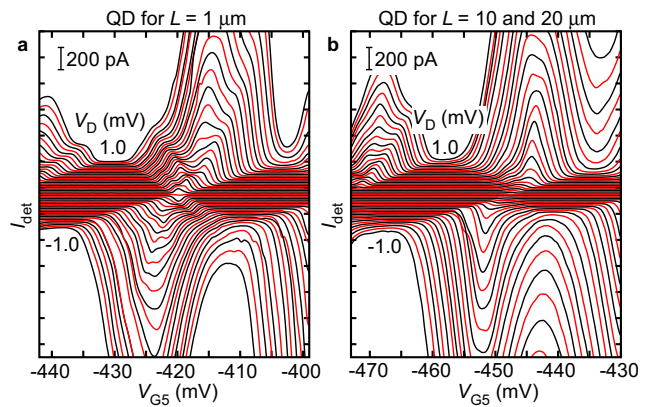}
\end{center}
\caption{\textbf{a} and \textbf{b} Coulomb diamond characteristics in $I_{%
\mathrm{det}}$ as a function of $V_{\mathrm{G5}}$ for various $V_{\mathrm{D}}
$ voltages from -1 mV to 1 mV in the step of 40 $\protect\mu $V for the QD
conditions at $L=$ 1 $\protect\mu $m in \textbf{a} and $L=$ 10 and 20 $%
\protect\mu $m in \textbf{b}.}
\end{figure}

\begin{figure*}[tbp]
\begin{center}
\includegraphics[width = 6.6 in]{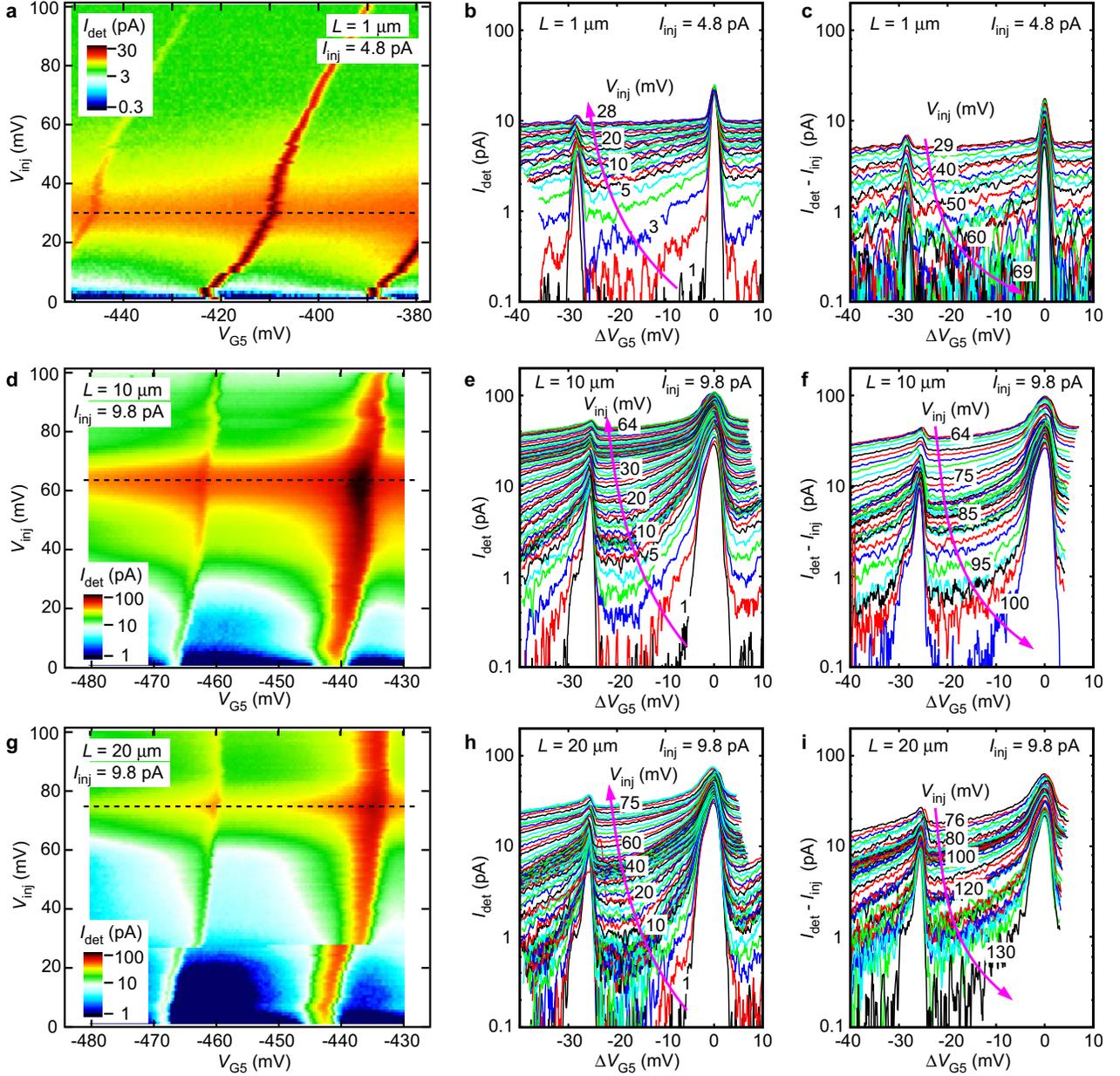}
\end{center}
\caption{QD detection current $I_{\mathrm{det}}$ for studying electron-hole
plasma near the Fermi energy. Panels a, b, and c for $L=$ 1 $\protect\mu $m,
panels d, e, and f for $L=$ 10 $\protect\mu $m, and panels g, h, and i for $%
L=$ 20 $\protect\mu $m are shown. Panels a, d, and g show color plots of $I_{%
\mathrm{det}}$ as a function of $V_{\mathrm{G5}}$ and $V_{\mathrm{inj}}$.
The dashed line shows the condition ($V_{\mathrm{inj}}^{\left( \mathrm{%
H\uparrow }\right) }$) for maximized plasma excitation. Panels b, e, and h
show $I_{\mathrm{det}}$ as a function of $\Delta V_{\mathrm{G5}}$ for
various $V_{\mathrm{inj}}<V_{\mathrm{inj}}^{\left( \mathrm{H\uparrow }%
\right) }$. Panels c, f, and i show $I_{\mathrm{det}}-I_{\mathrm{inj}}$ as a
function of $\Delta V_{\mathrm{G5}}$ for various $V_{\mathrm{inj}}>V_{%
\mathrm{inj}}^{\left( \mathrm{H\uparrow }\right) }$. }
\end{figure*}

\bigskip 

\textbf{SN2: QD detection}

Coulomb diamond characteristics of the QD in the absence of hot-electron
injection are shown in Figure 11a for the measurement at $L$ = 1 $\mu $m and
Fig. 11b for $L$ = 10 and 20 $\mu $m. The charging energy $U\simeq $ 1.5
meV, typical energy spacing $\Delta \simeq $ 0.2 meV, and the lever arm
factor $\alpha \simeq 0.044e$ to obtain a change in the electrochemical
potential $\Delta \mu _{N}=\alpha \Delta V_{\mathrm{G5}}$ for a change $%
\Delta V_{\mathrm{G5}}$ in $V_{\mathrm{G5}}$ were obtained from the data.

The electron-hole plasma under hot-electron injection is analyzed with the
QD, as shown in Figure 12. As shown in panels a, d, and g for $L=$ 1, 10,
and 20 $\mu $m, respectively, the positions of the CB peaks shift to higher $%
V_{\mathrm{G5}}$ side with increasing $V_{\mathrm{inj}}$, which can be
understood with the electrostatic effects from $V_{\mathrm{inj}}$ and $V_{%
\mathrm{Ginj}}$. While the effect is largest in Figure 12a for the shortest $%
L=$ 1 $\mu $m, the width and height of the CB peaks do not change so much,
indicating that the QD is well isolated from the IPC. Finite current in the
CB regions appears with increasing $V_{\mathrm{inj}}$ for a given $I_{%
\mathrm{inj}}$. This current is maximized at $V_{\mathrm{inj}}=V_{\mathrm{inj%
}}^{\left( \mathrm{H\uparrow }\right) }$, where the hot spot $x_{\mathrm{%
H\uparrow }}$ of the spin-up channel coincides with the QD position. The
condition $V_{\mathrm{inj}}^{\left( \mathrm{H\uparrow }\right) }$ increases
with increasing $L$ ($V_{\mathrm{inj}}^{\left( \mathrm{H\uparrow }\right) }=$
28, 64, and 75 mV for $L=$ 1, 10, and 20 $\mu $m, respectively), from which
the hot-electron trajectory is investigated in the main paper.

The non-thermal state after equilibration at $x=L>x_{\mathrm{H\downarrow }}$
is studied with the data at $V_{\mathrm{inj}}<V_{\mathrm{inj}}^{\left( 
\mathrm{H\uparrow }\right) }$ in Figure 12b, e, and h. In these
semi-logarithmic plots, the broad current profiles in the CB regions and the
sharp CB peaks coexist in the whole range. The data suggests the coexistence
of warm electrons ($T_{1}$) and cold electrons ($T_{0}$ close to the base
temperature), which justifies the binary Fermi-Dirac distribution function
as a non-thermal metastable state in the Tomonaga-Luttinger (TL) liquid.

The non-equilibrium state in the course of e-e scattering at $x=L<x_{\mathrm{%
H\uparrow }}$ was analyzed with the data at $V_{\mathrm{inj}}>V_{\mathrm{inj}%
}^{\left( \mathrm{H\uparrow }\right) }$ in Figure 12c, f and i. As the
injected hot electrons pass over the QD and are absorbed by the ammeter, $I_{%
\mathrm{det}}-I_{\mathrm{inj}}$ should be evaluated. The excess current in
the CB region decreases with increasing $V_{\mathrm{inj}}$, which suggests
suppression of e-e scattering for hot electrons at higher energies. This
suppression is discussed with momentum conservation in the main paper.

\bigskip 

\textbf{Data availability}

The data and analysis used in this work are available from the corresponding
author upon reasonable request.

\bigskip

\textbf{Acknowledgements}

We thank Taichi Hirasawa for preliminary measurements in the early stage of
the study. This study was supported by the Grants-in-Aid for Scientific
Research (KAKENHI JP19H05603) and the Nanotechnology Platform Program of the
Ministry of Education, Culture, Sports, Science and Technology, Japan.

\bigskip

\textbf{Author contributions}

T.F. designed and supervised this study. T.A. and K.M. grew the wafer. Y.S.
fabricated the device. K.S. performed the measurement and analysis with help
from T.H. and T.F. K.S. and T.F. wrote the manuscript. All authors discussed
the results and commented on the manuscript.

\bigskip

\textbf{Competing interests}

The authors declare no competing interests.

\bigskip

\textbf{Correspondence} and requests for materials should be addressed to
T.F.

\bigskip 

The final version of this paper was published in Communications Physics. 6,
103 (2023).

\end{document}